\documentclass[aps,pre,twocolumn,showpacs,floatfix]{revtex4}
\usepackage[dvips]{graphicx}
\usepackage{bm,amsmath}

\begin{document}
\title{A numerical technique for studying
topological effects on the thermal properties of knotted polymer rings}
\author{Yani Zhao}
\email{yanizhao@fermi.fiz.univ.szczecin.pl}
\author{Franco Ferrari}
\email{ferrari@fermi.fiz.univ.szczecin.pl}
\affiliation{CASA* and Institute of Physics, University of Szczecin,
  Szczecin, Poland} 
\date{\today}
\begin{abstract}
The topological effects on the thermal properties of
several knot configurations are investigated using  Monte Carlo
simulations.  
In order to check if  the topology of the knots
is preserved
 during the
thermal fluctuations
we propose a method  that allows very fast calculations and can be
easily applied 
to arbitrarily complex knots.
As an application, the specific energy and heat capacity of the
trefoil, the figure-eight  and the $8_1$ knots are calculated at
different temperatures and 
for different lengths. Short-range repulsive interactions between the
monomers are assumed. The knots configurations are generated on a
three-dimensional 
cubic lattice and sampled by means of the Wang-Landau
algorithm and of the pivot method. The obtained results show that the
topological effects play a key 
role for short-length polymers. Three temperature regimes of
the growth rate of the internal energy  of the system are distinguished. 
\end{abstract}
\maketitle
\clearpage
\section{Introduction}
Topological polymers are a fascinating subject by themselves.
Moreover, the interest in this topic is motivated by possible
industrial applications, for example because knots in polymer materials are able to
influence the viscoelastic properties \cite{grosberg}, and due to their relevance in biology and
biochemistry, see for instance \cite{dna,katritch96,katritch97,krasnow, laurie,cieplak,liu,wasserman,sumners}.
Polymer rings may be found in nature or in artificial materials
in the form of knots that can be additionally linked together.
In this work we explore the case of the statistical mechanics of a
single polymer knot. 
Various topological configurations are taken into account: The unknot,
denoted according to the Alexander-Briggs notation $0_1$,
the trefoil $3_1$, the figure-eight $4_1$ and the knot $8_1$.
In particular, we study the thermal properties of the above knots by
computing numerically their internal energy and heat capacities.
The most important problem of numerical simulations of polymer knots
is to preserve their topology while their configurations are
fluctuating randomly. To this purpose, several methods have been
developed
\cite{vologodski,orlandini,kurt,mehran,pp2,metzler,muthu}. Most of
them, are
based on the Alexander or Jones polynomials, which are rather powerful
topological invariants able to distinguish with a very high degree of accuracy the
different topological configurations. The main drawback
of these 
polynomials is that their calculation is very expensive and time
consuming from the computational point of view.

Here we propose a new strategy that
consists in the following. First, we apply
the pivot algorithm to change a given knot
configuration. 
Starting from a seed knot, for instance those given in
ref.~\cite{scharein},
we change it at randomly chosen points by applying the pivot algorithm \cite{pivot}.
After each pivot move, the difference between the old and new configurations forms
a closed loop. Around it, an arbitrary surface is stretched, whose
boundary is the closed loop itself. The criterion to reject changes that destroy the
topology of the knot is the presence or not of lines of the old knot that
cross such surface. If these lines are
present, the trial pivot move is rejected. This combination of 
pivot algorithm and excluded area (PAEA) provides an efficient
and fast way to preserve the topology that can be applied to any
knot configuration, independently of its complexity. For pivot moves
involving a small number of segments the method becomes exact.
This technique may be employed in the study of the thermal and mechanical
properties of polymer knots. Here we have limited ourselves to the
computation of the internal energy and heat capacity in the case in
which the monomer interactions are of very short range, but the
extension to any other interactions or to the calculations of more
complicated physical quantities is possible.
The configurations of the unknot, trefoil, figure-eight and $8_1$ knot
are modeled on a cubic lattice. The sampling
of the canonical ensemble is achieved by using the Wang-Landau
algorithm \cite{wanglaudau} at different temperatures.

The rest of this paper is organized as follows. In Sec.~II, we describe
the sampling and simulation methods of the polymer knots. 
Our strategy  to distinguish possible violations of the topology
of a given knot after a random change of its trajectory is explained.
The
calculation of the specific
energy and the specific heat
capacity for various polymer topologies and for different polymer lengths is
presented in section~III. In section~IV we draw our conclusions and 
discuss the open problems together with new possible directions of our
research.\\

\section{Model and Computational method}

\begin{figure}
\begin{center}
\includegraphics[width=3in]{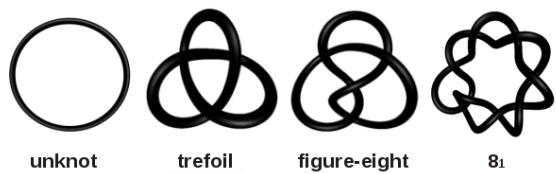}
\caption{\label{bentstraightwirefigure}
In this work different knot types will be
  studied, namely the unknot, the trefoil, the figure-eight and the
  $8_1$ knot.} 
\end{center}
\end{figure}

\begin{figure}
\includegraphics[width=3in]{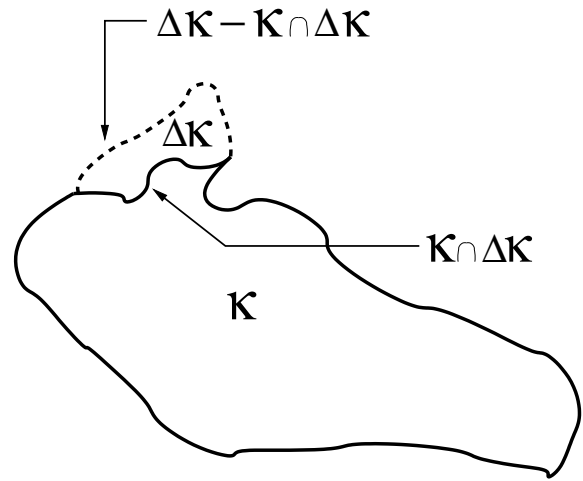}
\caption{This figure represents in a schematic way the meaning of the
  symbols used to denote the element ${\cal K}\bigcap\Delta {\cal K}$ of
$\Delta {\cal K}$ shared with the knot $\cal K$ and the element 
$\Delta{\cal K}- {\cal K}\bigcap\Delta {\cal K}$ 
of $\Delta {\cal K}$
which is not shared with
$\cal K$.\label{kdkloops} }
\end{figure}

In the current work, we will study the topological effects on the
thermal properties of polymer rings. Four types of polymers will be
considered on a 3-dimensional cubic lattice, namely the unknot ${\cal
  U}$, the trefoil ${\cal T}$, the figure-eight ${\cal F}$ and the
knot $8_1$, see Fig.~\ref{bentstraightwirefigure}. 
A general knot will be denoted with the symbol ${\cal K}$.
To simulate the trivial polymer ring ${\cal U}$, we generate a loop
starting from a 
random point and performing a self-avoiding walk on the cubic
lattice. The walk is stopped when it crosses itself, forming a
loop. We generate many loops of this kind and we retain only those
with a given length $L$. For nontrivial topologies, we
choose a different strategy. We begin from a seed conformation and
then we act repeatedly on it with the pivot method~\cite{pivot}  to get
random conformations. The elements of the loop that will be changed by
the pivot move have a fixed number of segments $N=4$ or $N=5$.  
The elements begin at a random point in the knot, let's say  $N_{0}$
and end up at the point number $N_0+4$, or $N_0+5$. $N_{0}$ can range
from 0 to $L-N$ ($L$ is the length of the polymer). We check the
topology of the configuration after every pivot move, to be sure that the
move will not change the topology of the given knot. Usually, this
check is achieved by
projecting the knot along some direction on
a two dimensional plane and the relative Alexander polynomial is
computed. This method has been pioneered in
refs.~\cite{vologodski}. Here we choose another strategy, namely on a
given knot ${\cal K}$ we perform a pivot move. In that way a new
knot ${\cal K}'$ is obtained. 
A schematic picture of the situation can be found in fig.~\ref{kdkloops}.
In order to verify that the topology of
the knot has not been 
changed by the pivot move, we check that no lines of ${\cal K}$ 
cross the area
spanned by the small loop $\Delta{\cal K}={\cal K}'-{\cal 
  K}$ including its border. Otherwise, this could imply
that a crossing of the lines of ${\cal K}$ has happened. 
The definition of the
area of $\Delta{\cal K}$ in a numerical simulation is a difficult
problem of computer graphics. However, if
the number of segments involved in the pivot moves is small, it is
possible to
classify all the  geometries  of $\Delta{\cal K}$ and to stretch
a suitable surface $\Sigma_{\Delta\cal K}$ around $\Delta{\cal K}$.
Next, these data, namely all  geometries of $\Delta{\cal K}$
and the coordinates of the relative surfaces  $\Sigma_{\Delta\cal K}$,
are supplied to the code, which will then be
able to detect the occurrence of possible lines of ${\cal K}$ crossing
$\Sigma_{\Delta\cal K}$ or the part of its border
 $\Delta{\cal K}-{\cal K}\bigcap \Delta{\cal K}$, see
fig.~\ref{kdkloops} for a schematic representation of what these
symbols mean.
If these lines are present, the trial pivot move is rejected.
Let us note that the strategy described above could be considered as
an implementation of the topological 
constraints based on the Gauss linking number modified to cope with
the present situation, in which the Gauss linking number becomes not
suitable because $\cal K$ and $\Delta\cal K$ share part of their
trajectories. While the Gauss linking number is a weak topological
number,
our method turns out to be 
very efficient when
applied to pivot modes with
segment lengths $N=4$ and $N=5$. In both cases the geometry of the
loop $\Delta\cal K$ is simple enough, that it is possible to analyze all
situations in which the crossings of lines could violate the topology of $\cal
K$ and to
provide an exact criterion
 for preserving this topology  without  increasing exceedingly the
 simulation time.  Only when $N>5$ it may occur 
that the trajectory of the old knot crosses the area
$\Sigma_{\Delta\cal K}$ an even number of times. 
In this case, due to limitations similar to those of the Gauss
linking number,
allowed moves can 
be rejected. If this happens very frequently,
the computational time may increase. However,
this is not a serious drawback because
the above procedure of detecting topology changes is
extremely time efficient. The necessary cpu time depends essentially
only on the number of segments $L$, independently on the complexity of
the knot.
In the following, we will restrict ourselves to the cases
$N=4$ and $N=5$. 
\\ 
In order to calculate the internal energy of the polymer per unit of
length $\langle E(\beta)\rangle/L$, 
where $\beta=T^{-1}$ denotes
the Boltzmann factor with $k_B=1$ (natural thermodynamic units) and
the heat capacity
\begin{equation}
 C(T) =\dfrac{1}{T^{2}}(\langle E(\beta)^{2}\rangle -(\langle E(\beta)\rangle)^2)\label{heatcapacity}
\end{equation}
 the density of states of polymers is needed. To that purpose, the
 Wang-Landau (WL) algorithm is used for sampling the polymer
 conformations. In the case of the trefoil, figure-eight and $8_1$ knot, before
 applying the WL algorithm, the system has been equilibrated, since we
 start from a given seed which is not in a random conformation.  The
 fluctuations of  
 the radius of gyration $R_G$ are employed to judge the degree of
 equilibration. These 
 fluctuations are computed according to the formula \cite{rq}: 
\begin{equation}
\Delta R_{G}^{2}(t)=\sqrt{\dfrac{1}{t} \sum_{\tau=1}^{t}(R_{G}^{2}(\tau)-\langle R_{G}^{2}\rangle_{t})^{2}},
\end{equation}
Here $t$ is the sweep time, $\langle R_{G}^{2}\rangle_{t}$ is the
average radius of gyration at time $t$. The equilibrating procedure is
stopped after the time dependent fluctuation of the radius of
gyration $\langle\Delta R_{G}^{2}\rangle$ oscillates around a
constant. 
To crosscheck the equilibration of the nontrivial knot configurations, we use
also another method, that consists in counting the number of sites in
which the direction of the chain is not changed  \cite{tk}. This
number is a random variable, but at
equilibrium its variance around the average is small.
The results obtained with both  techniques, that based on the
fluctuation of the gyration radius 
and that based on the changes of direction of the segments, are in
agreement. 
\begin{figure}
\centering
\includegraphics[width=3in]{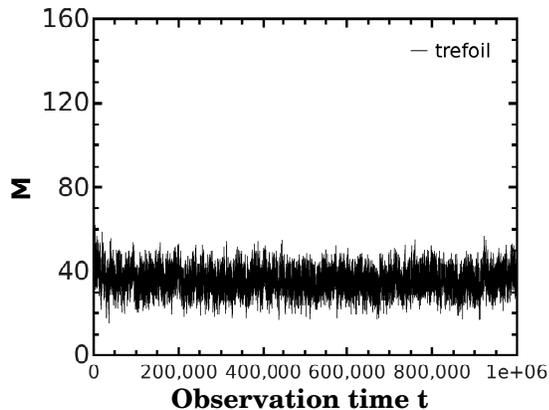}
\caption{\label{equilibration} $M$ is the number of sites of the trefoil in which the direction of the chain is not changed.}
\end{figure}
Fig.~\ref{equilibration}, which is related to the latter method, shows that the system
equilibrates after a few tens of thousands of pivot moves in the case
of the trefoil ${\cal T}$. Similar considerations are valid for the
other knots considered in this work. In total the system is subjected
to one   
million of pivot moves before treating it with the WL algorithm in
order to compute the energy $\langle E(\beta)\rangle$ and the heat
capacity $ C(T)$.

\begin{figure}
\begin{center}
\includegraphics[width=3in]{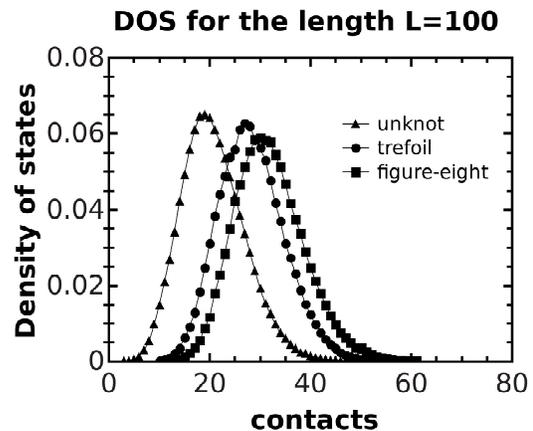}
\caption{ \label{dos} The density of states is obtained
  by  WL
  algorithm. The results for the unknot (triangles), the trefoil
  (circles) and the figure-eight (rectangles)
  are represented.} 
\end{center}
\end{figure}

Now we briefly introduce the WL algorithm. This method is considered as
a self-adjusting procedure for obtaining the density of states. 
In the present context the states are distinguished by the number $i$ of
closest contacts between the polymer monomers, where $i$ takes
positive integer values.
Following
a previous work \cite{pn}, all  initial values of
the density of states are firstly set to be equal to one,
i.~e. $\Omega_{i}=1$ for all the states. Then the pivot method is used to
visit the next trial state $i'$. The energy of the trial state will be
denoted by $\varepsilon m$, where $m$ is the number of the closest
contacts of the monomers, while $\varepsilon$ is the interaction energy of each
monomer pair. $\varepsilon >0$ and $\varepsilon <0$ correspond to
repulsive and attractive case respectively. We assume in the following
that $\varepsilon >0$. The probability of translating
from the state $i$ to $i'$ is 
\begin{equation}
p(i-i') =min[1,\dfrac{\Omega_i}{\Omega_{i'}}]
\end{equation}
Once an energy state is visited, the corresponding density of states
of this state will be updated by multiplying a modification
factor~\cite{wanglaudau}. After the $n-$th sweep, the modification
factor $f_i$
is reduced according to the equation $f_{i+1}=\sqrt{f_{i}}$. The above
process is repeated until the modification factor reaches the predefined limit
$f_{final}=\exp(10^{-8})$~\cite{wanglaudau}. 
Of course, at each step the preservation of the knot topology is
checked with the method explained before.
We show as an example the obtained densities of
states for the unknot, the trefoil and the figure-eight of length $L=100$
in fig.~\ref{dos}.\\ 
The obtained distribution of density of energy states can now be used to
calculate the canonical averages of the observables according to
standard relations. As already mentioned, the observables studied in
this paper are the
internal energy of the canonical system and the heat capacity as a
function of the temperature. The
expression of the internal energy is given by:
\begin{equation}
\langle E(\beta)\rangle  =\dfrac{\sum_{m}m\varepsilon e^{-\beta m
    \varepsilon}\Omega_{m}}{\sum_{m} e^{-\beta m
    \varepsilon}\Omega_{m}} 
\label{energy}
\end{equation}
where $m$ is the number of the closest contacts of polymers
and $\Omega(E)$ is the density of energy states computed by means of
the WL algorithm. 
The formula of the heat
capacity has been already provided in Eq.~(\ref{heatcapacity}).

The computational method has been checked using the 
trivial knot $\cal U$ as a test. We could repeat the previous
results of \cite{pnring} in the case of chains with lengths equal to
12, 30 and 50. 
\begin{figure*}
\begin{center}
{\includegraphics[width=0.5\textwidth]{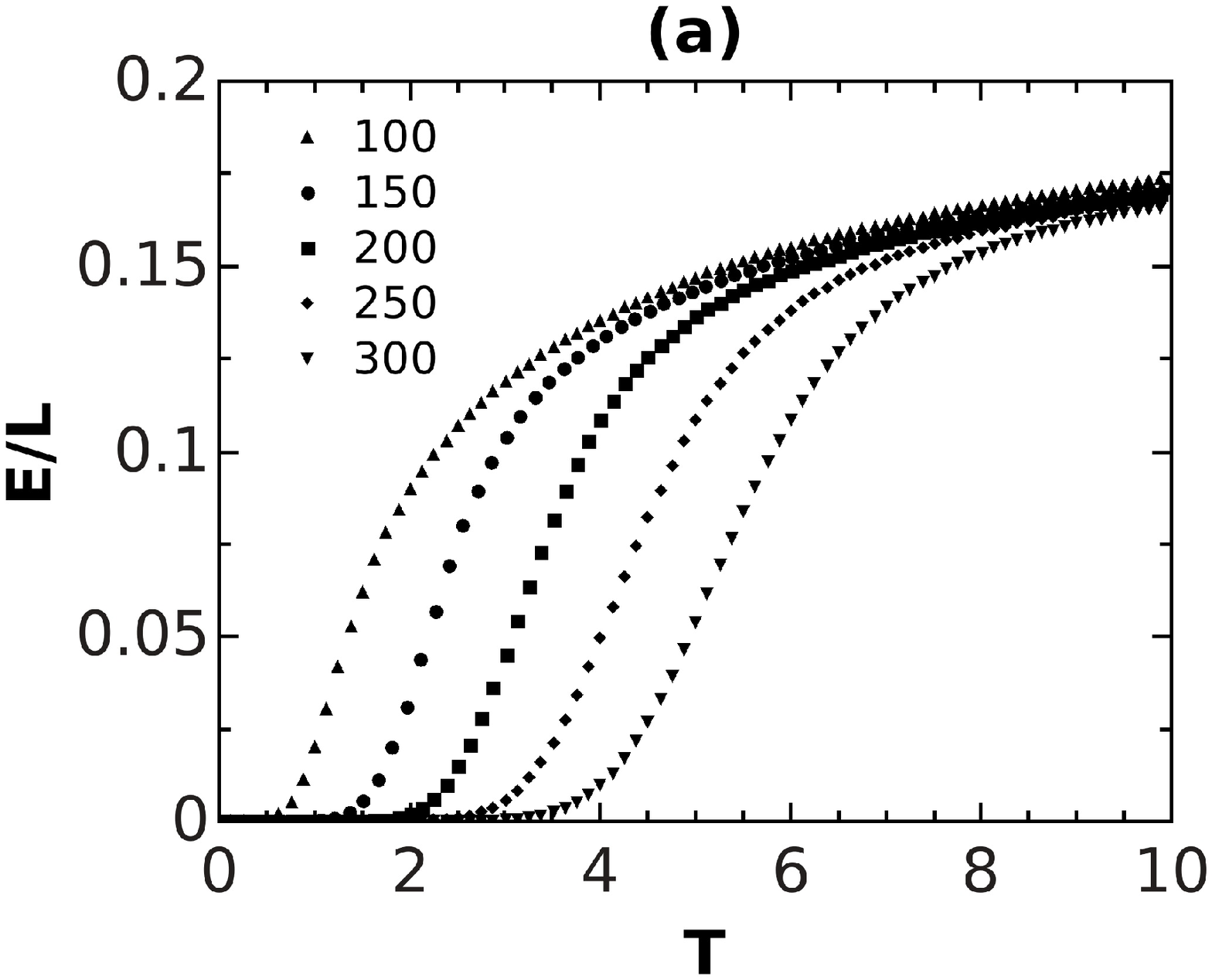}\includegraphics[width=0.5\textwidth]{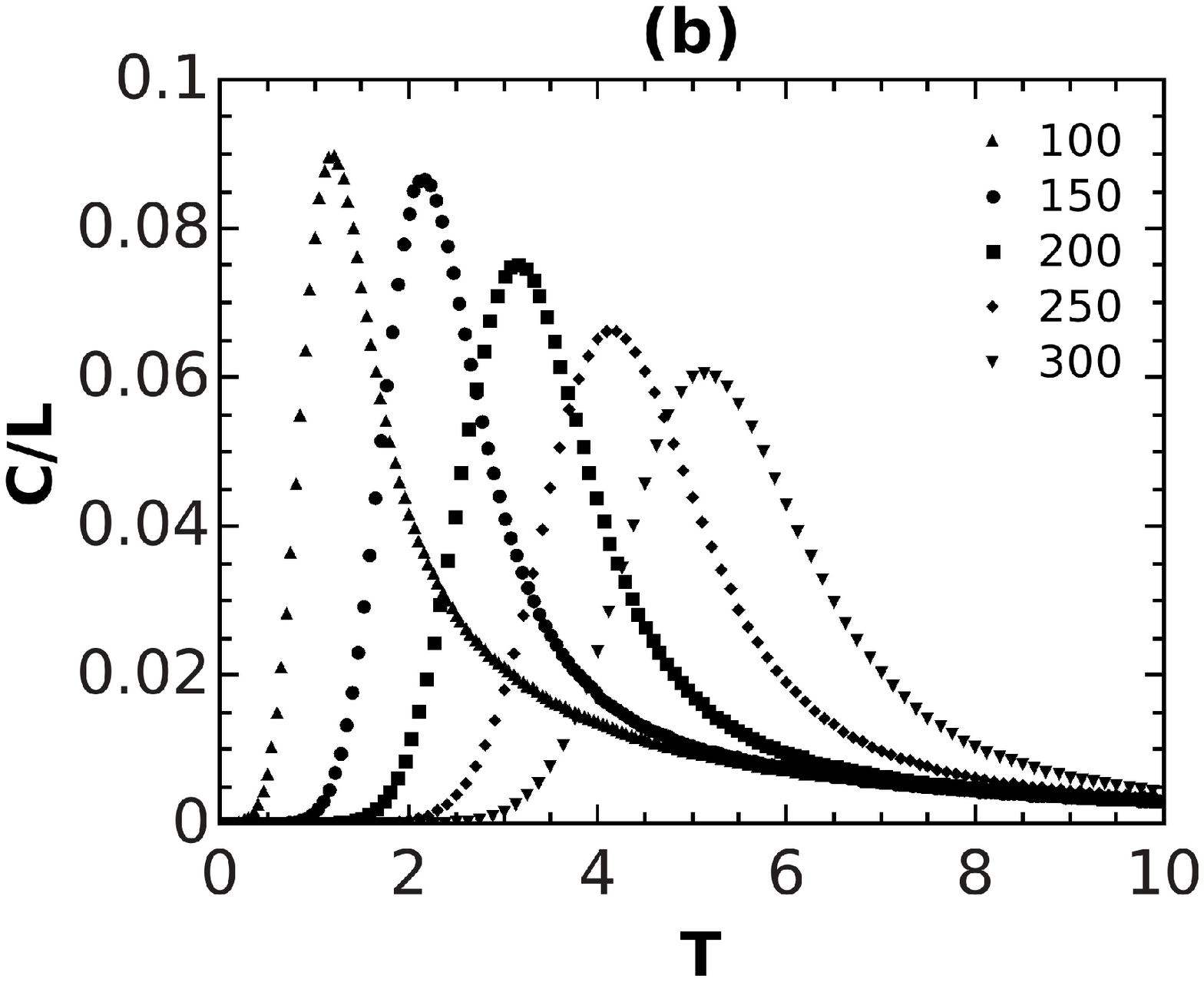}}
\caption{\label{geo} Specific
  energy and heat capacities for the unknot polymer as functions of
  the normalized temperature ${\mathbf T}=\frac T\varepsilon$. The
  polymer length can take the values 
   $L=100 \mbox{ (upper triangles)}, 150 \mbox{ (circles)}, 200 \mbox{
    (rectangles)}, 250 \mbox{ (diamonds)}, 300 \mbox{
    (down-triangles)}$, where $L$ denotes the number of
  steps. (a) Plot of the specific
  energy (in $\varepsilon$-units) of the unknot; (b) Plot of the specific heat
  capacity (in $\varepsilon$-units) of the unknot.}
\end{center} 
\end{figure*}
\section{results and discussion}
In this section we study the thermal properties of the
trefoil and figure-eight with  lengths in the range of $100-300$
segments for the case
$\varepsilon >0$. The used algorithm is quite fast. Depending on the length, the typical simulation times 
vary between 15 minutes to 1 hour on a dedicated pc. It can be applied
to much more complicated knots. As an example, we have analysed here
also the case of the $8_1$ knot.

\begin{figure*}
\centering
{\includegraphics[width=0.5\textwidth]{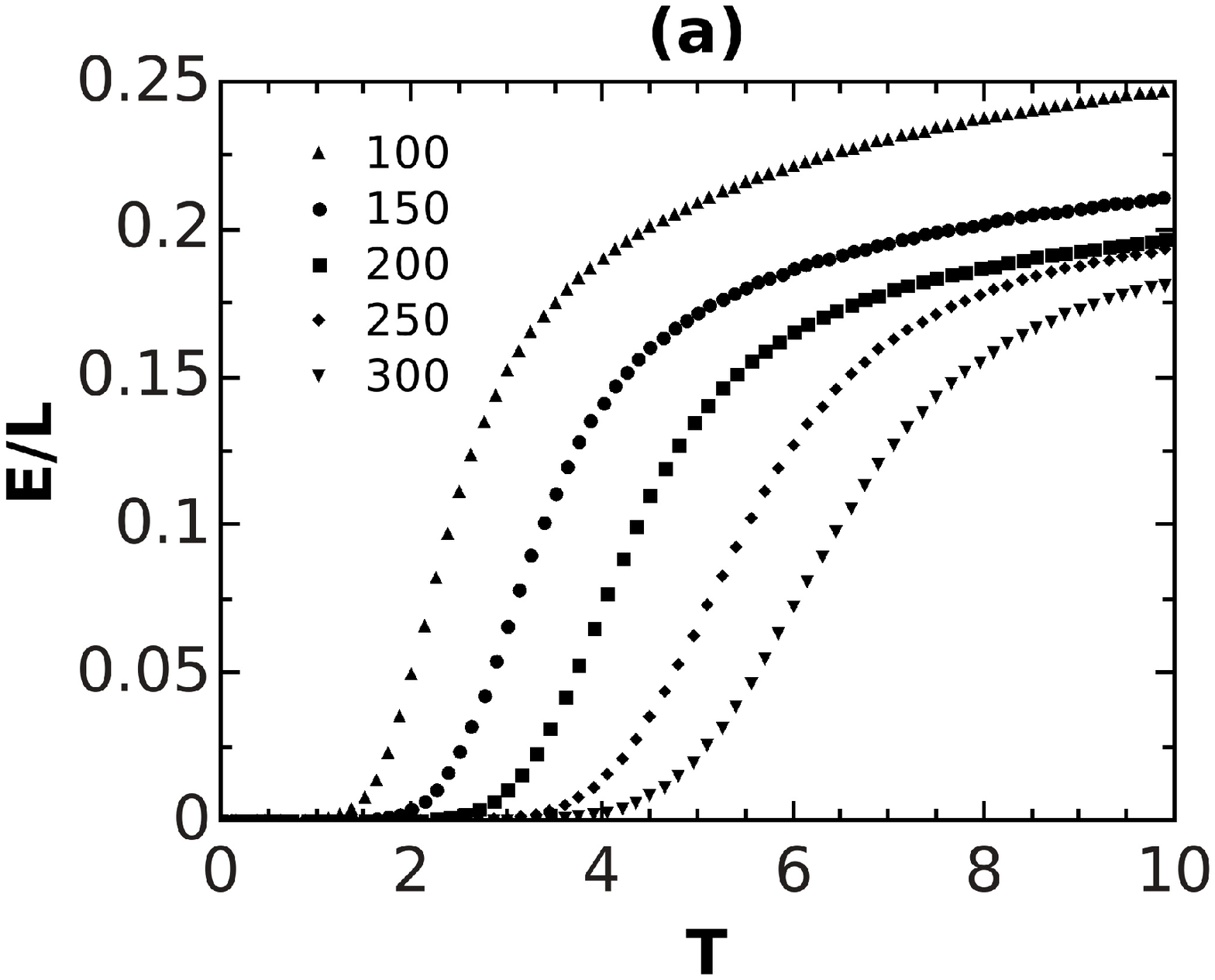}\includegraphics[width=0.5\textwidth]{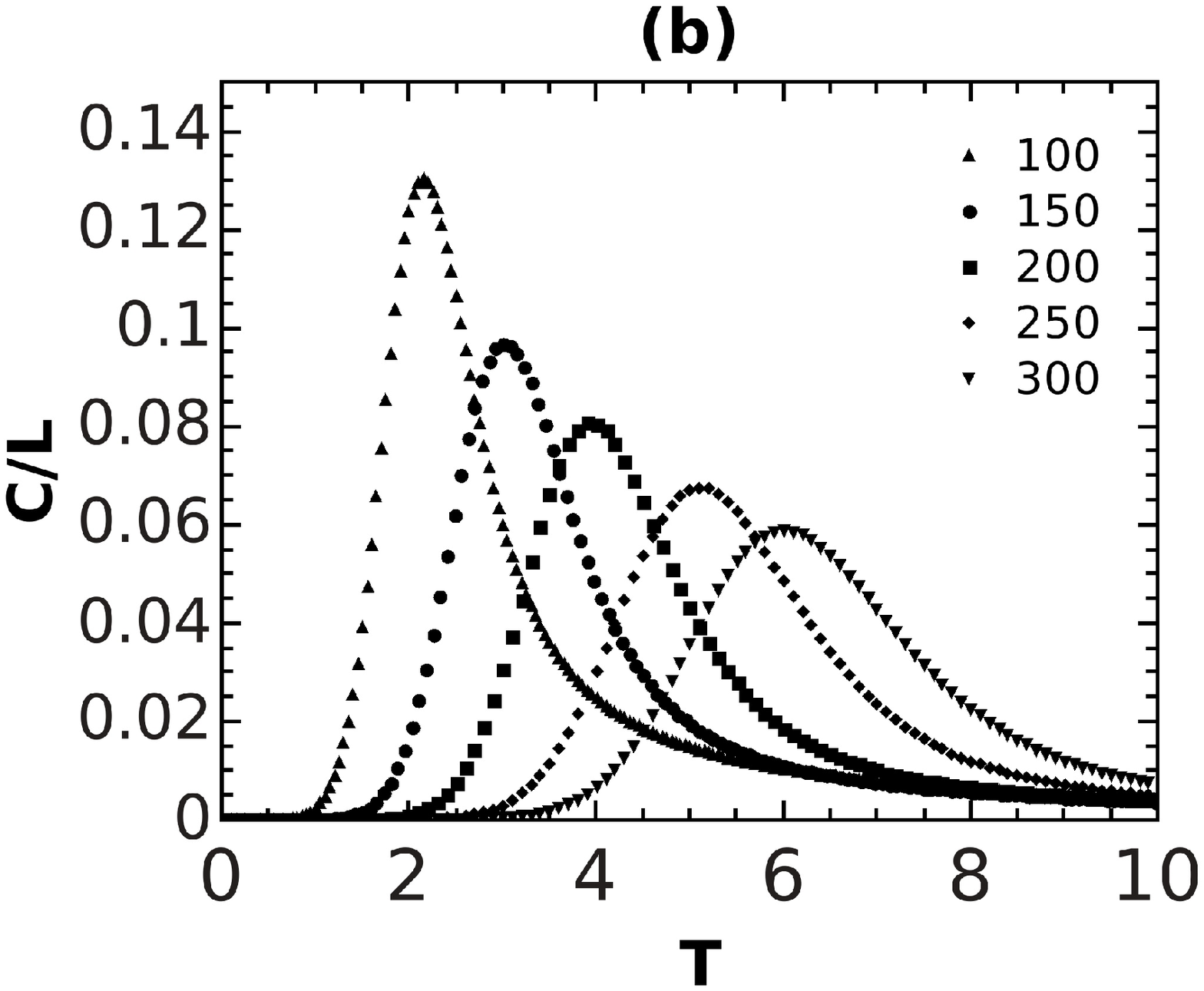}}
\caption{
\label{geo2} Specific
  energy and heat capacities for the trefoil as functions of
  the normalized temperature $\mathbf T=\frac T\varepsilon$. Separated
  plots are provided for trefoils with 
  $L=100 \mbox{ (upper triangles)}, 150 \mbox{ (circles)}, 200 \mbox{
    (rectangles)}, 250 \mbox{ (diamonds)}, 300 \mbox{
    (down-triangles)}$. (a) Plot of the specific 
  energy (in $\varepsilon$-units) of the trefoil; (b) Plot of the specific heat
  capacity (in $\varepsilon$-units) of the trefoil.}
\end{figure*}

\begin{figure*}
\begin{center}
{\includegraphics[width=0.5\textwidth]{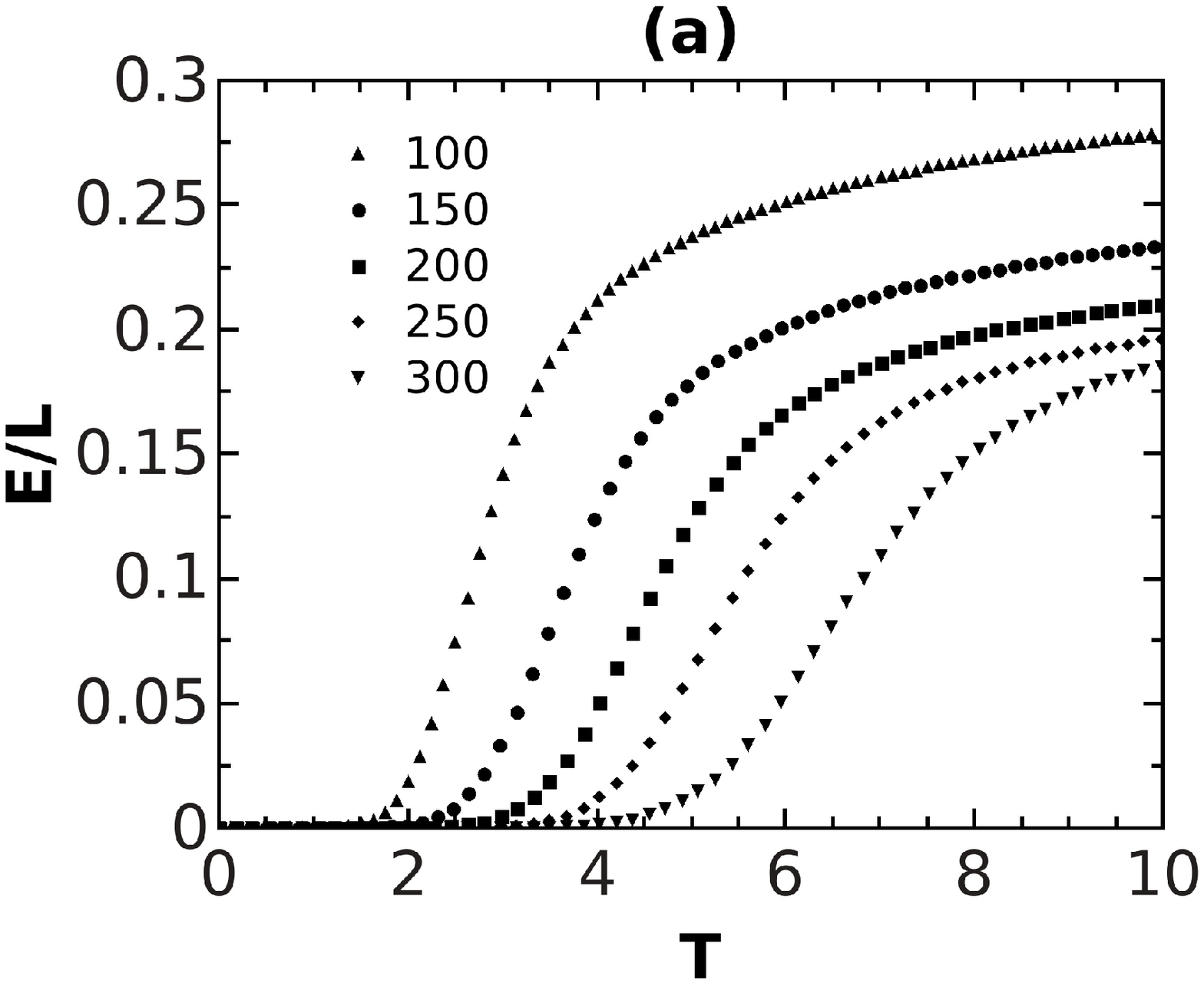}\includegraphics[width=0.5\textwidth]{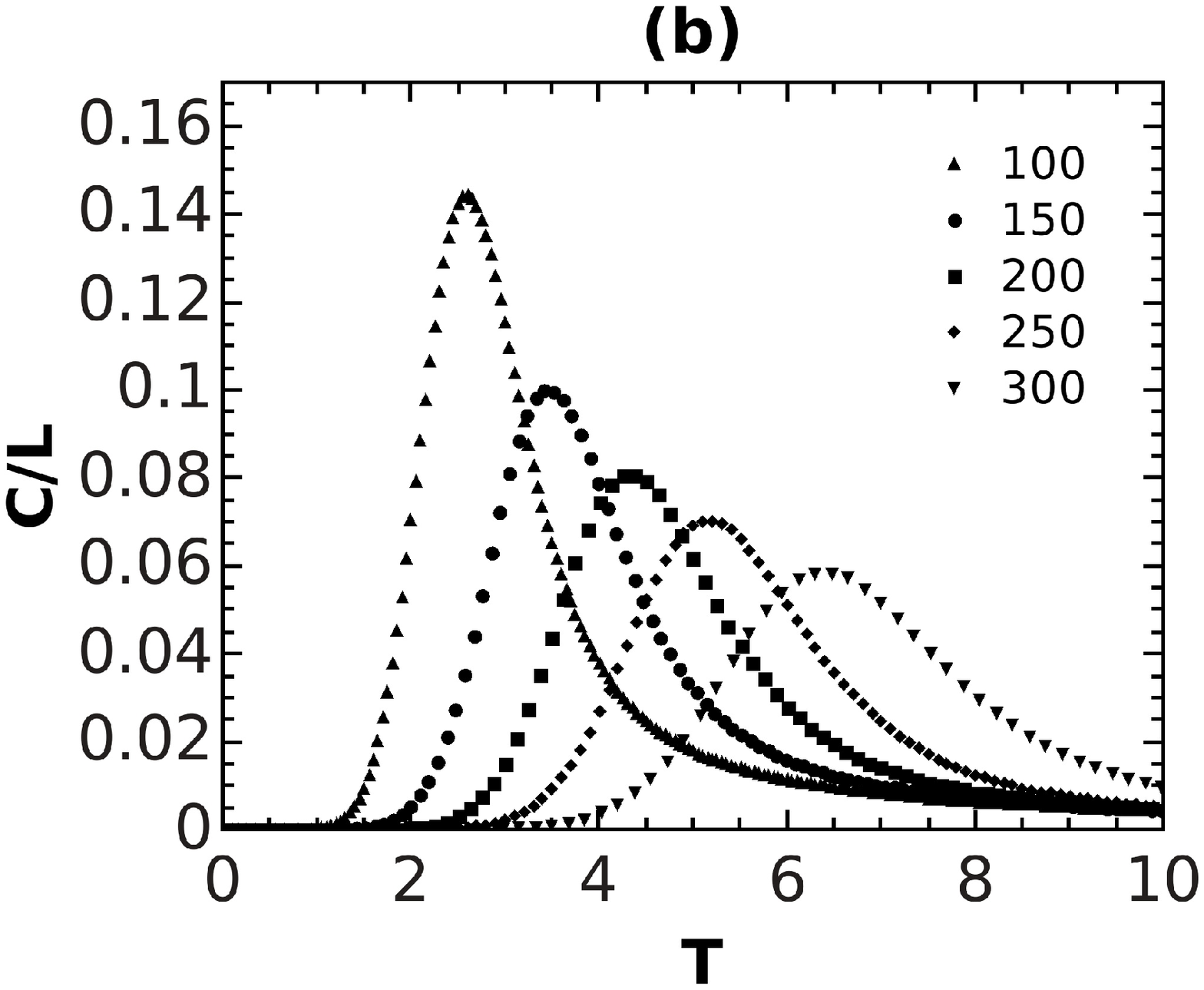}}
\caption{\label{geo3} Specific
  energy and heat capacities for the figure-eight as functions of
  ${\mathbf T=\frac T\varepsilon}$. The lengths are  $L=100 \mbox{
    (upper triangles)}, 150 \mbox{ 
    (circles)}, 200 \mbox{ 
    (rectangles)}, 250 \mbox{ (diamonds)}, 300 \mbox{
    (down-triangles)}$ steps. (a) 
  Plot of the specific energy (in $\varepsilon$-units) and (b)
  plot of the specific heat capacity (in $\varepsilon$-units) of the
  figure-eight knot.} 
\end{center}
\end{figure*}

The results concerning the trefoil and  figure-eight are presented in
fig.~\ref{geo2} and fig.~\ref{geo3} respectively. 
For comparison, fig.~\ref{geo} represents the results for $\cal U$
with 
 lengths ranging from $100$ to $300$. 
It is found that the growth of the specific
energies $\langle E(\beta)\rangle/L$ in  
fig.~\ref{geo2}(a) and fig.~\ref{geo3}(a) is characterised by three regions both for
the trefoil and the figure-eight. At very low temperatures
the energy increase is practically zero. The reason is that the
temperatures are too low to allow contacts between the monomers. When
the energy of the system grows -- let's remember that we are working in
energy units, so that we can talk about temperatures and energies
interchangeably -- the ``first energy state'' can be reached. This
consists in the case in which at least two monomers are in contact. 
For instance, we expect that
when the chain has length $L=100$, the first energy state occurs in
figs.~\ref{geo2} and \ref{geo3} at some ``critical'' temperature
$T_1\sim 1$, while when $L=200$ this temperature should be given by $T_1\sim
2$. These values of $T_1$ slightly increase with increasing topological
complexity.
In general, this proportionality temperature/length is respected
in figs.~\ref{geo2} and \ref{geo3}.
After  the first energy state is reached, the specific energy grows
rapidly as a 
function of the temperature until saturation is reached after a given
temperature $T_2$ and the energy
increase becomes moderate. 
The division into three regions can be expected by looking at
the Boltzmann factor $e^{-\beta m \varepsilon}$ appearing in the expression
of the energy given in Eq.~(\ref{energy}):
\begin{description}
\item{i)} In the range $T<T_{1}$, the
factor $e^{-\beta m \varepsilon}$ goes rapidly to $0$, because low
temperatures cause 
few states to be occupied in this situation, so we get $\langle
E\rangle \rightarrow 0$. 
\item{
ii)} When $T_{1}<T<T_{2}$, the factor
$e^{-\beta m \varepsilon}$ changes sensitively depending on
$\beta=T^{-1}$. In this regime more and more states of the given
  system are getting excited
through heat absorption and $\langle E\rangle $  increases 
with the increase of the temperature $T$ as expected.
\item{ iii)} Once $T>T_{2}$,
$\langle E\rangle $ increases slower due to the fact that most states
  have been already
occupied in the range $T_{1}<T<T_{2}$. Just  few states are left that can be
excited in this situation. One could notice that the factor $e^{-\beta
  m \varepsilon}$ in Eq.~(\ref{energy}) increases slowly at high
temperatures. When $T \rightarrow \infty$, we have
$e^{-\beta m \varepsilon}\rightarrow 1$.
\end{description}
It can also be found that in fig.~\ref{geo2}(b) and
fig.~\ref{geo3}(b), the heat capacity has a peak in the range of
temperature $T_{1}<T<T_{2}$. The appearance of this peak together with
the sigmoidal behavior of the specific energy points out at an ongoing
two-state phase transition according to \cite{cooper}. The presence of
these
pseudo-transitions related to the lattice geometry have been clarified
in ref.~\cite{wustlandau,vogel}.
We would also like to show that the heat capacity and energy are
affected not only by the temperature, but also
by  the polymer length $L$ and its topological
conformation. First of all, the specific energy of the chain decreases with
increasing values of $L$. 
For instance, the quantity $\langle E(\beta)\rangle/L$ for
 $L=100$ is larger than that of longer chains as it is illustrated in fig.~\ref{geo2} and
fig.~\ref{geo3}. This is because longer polymers in the presence of
short-range repulsive interactions prefer states
 which are diluted comparing with the short ones. Probably for this reason,
 in proteins knots are localized in regions with a very small
 number of residues~\cite{dna}.
The total internal energy, instead, decreases with decreasing polymer lengths.
This is due to the fact that longer polymers
can accommodate a larger number of contacts. 
Let us note that the peak temperatures of the
heat capacity are shifting
with the length $L$. This is expected since
we are computing the specific heat capacity. If we computed the total
heat capacity of the system, these peaks would occur at the same temperature.

Concerning the topological effects, we observe that both the energy and
the heat capacity increase with increasing knot complexity. This can
be seen by comparing the plots for various knots of the same length in
figs.~\ref{geo}, \ref{geo2} and \ref{geo3}.
On the other side, the topological effects are stronger
 when polymers are shorter. This is shown in 
Table~\ref{table:nonlin} in which the energy differences for
polymers with the same length but different topology are listed.
In the table $d_{UT}$ is the difference between the
unknot and the trefoil defined by
$d_{\cal UT}\equiv\frac{\langle E_{\cal U}\rangle -\langle
E_{\cal T}\rangle}L $, while $d_{\cal TF}$
denotes the difference between the trefoil and figure-eight. When the
length ranges from $100$ to $300$, $d_{\cal UT}$  varies from
$0.073$ to $0.015$ and $d_{\cal TF}$ from $0.032$ to $0.004$
at $T=10\varepsilon$. The heat capacity exhibits a similar
behavior. It is thus 
pretty clear that 
the topological effects become less important when the length
is increasing. This is similar to the so called 'size effect' in
nanomaterials~\cite{yani}.
Topological
properties will have a lesser influence in the case of diluted
monomers.  We expect that topological effects will disappear if the
polymer is long
enough. \\ 

\begin{figure*}
\centering
{\includegraphics[width=0.5\textwidth]{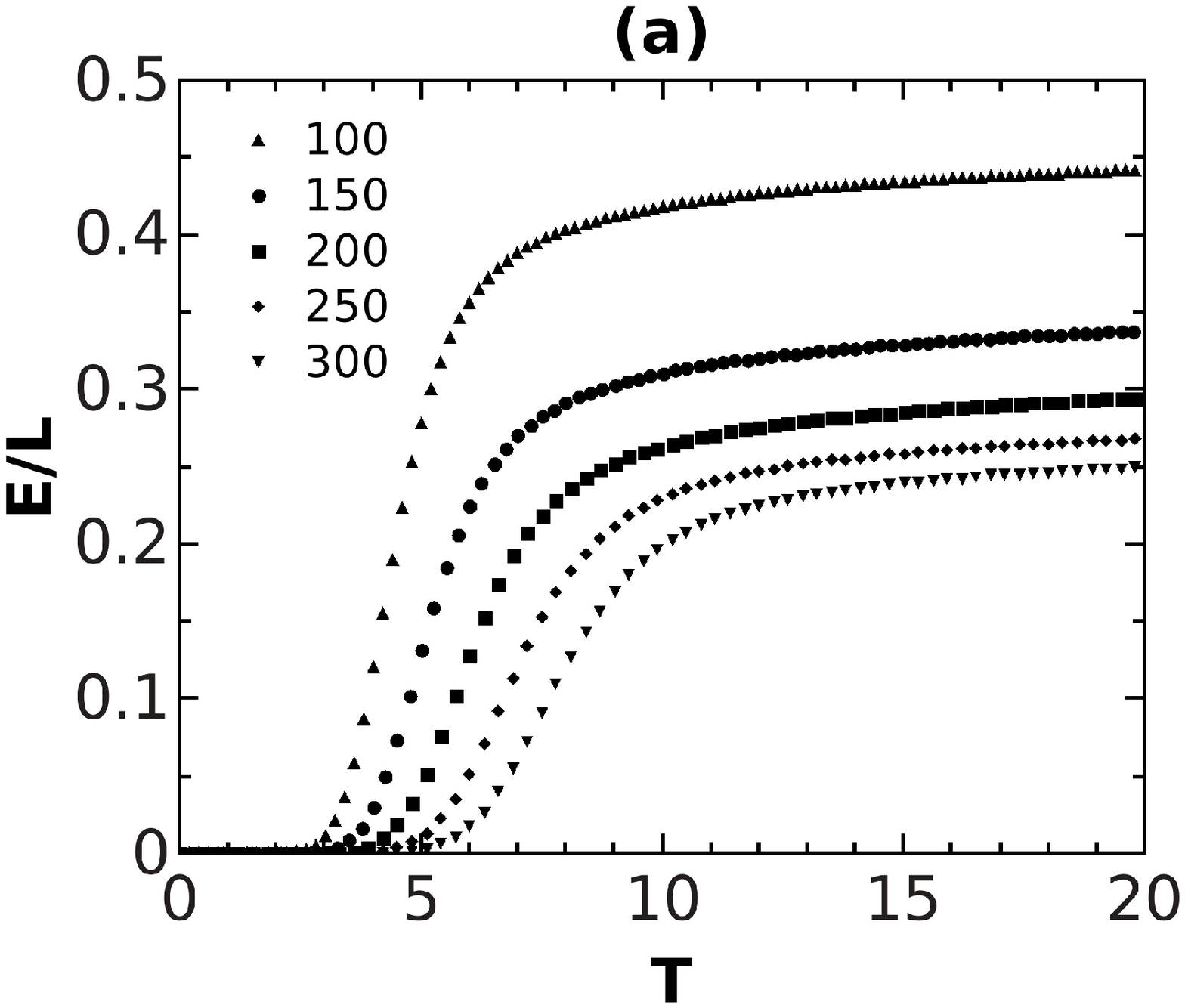}\includegraphics[width=0.5\textwidth]{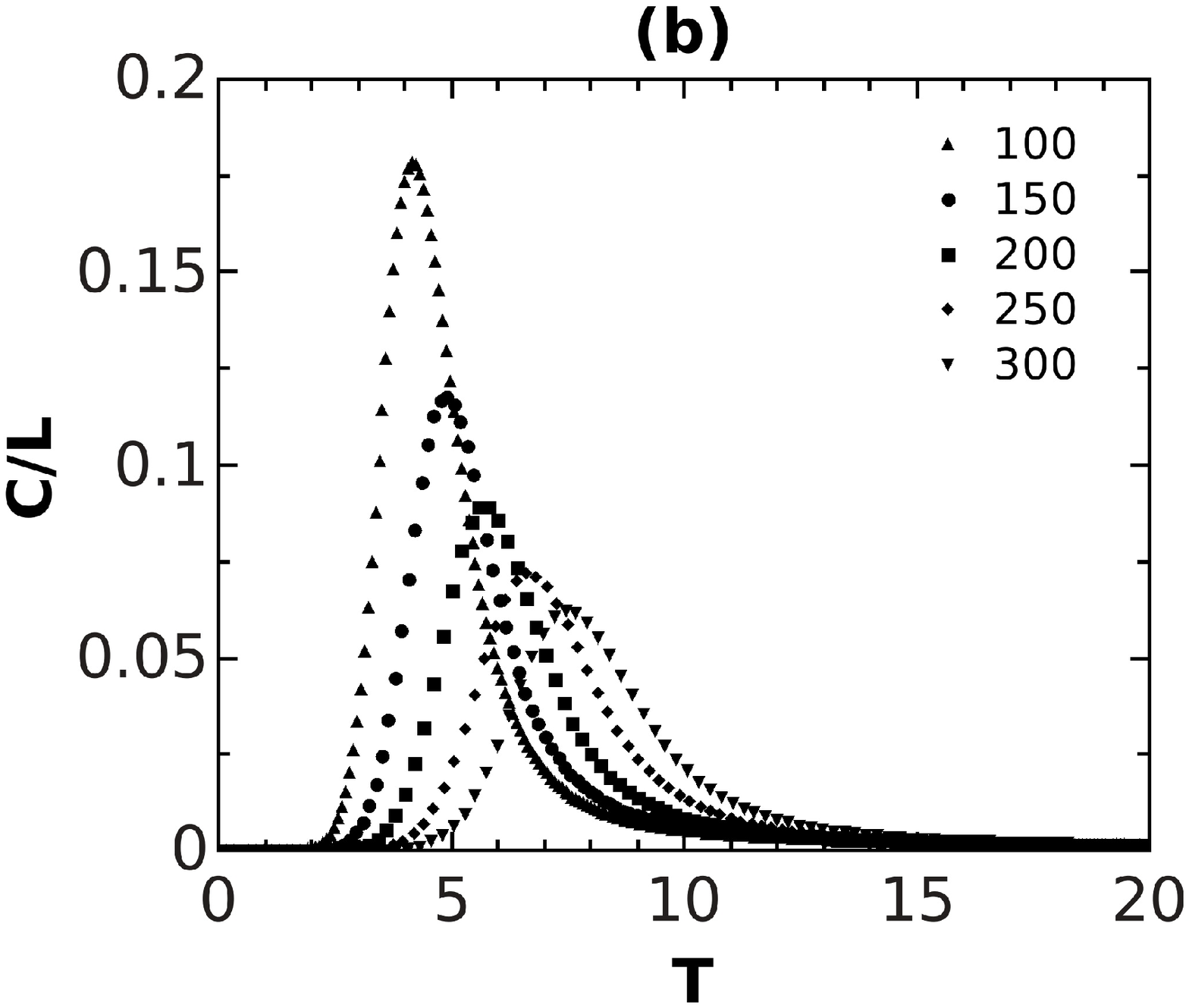}}
\caption{\label{geo4} Specific
  energy and heat capacities for ${\cal K}=8_1$ as functions of
  $\mathbf T=\frac T\varepsilon$. The studied lengths are
   $L=100 \mbox{ (upper triangles)}, 150 \mbox{ (circles)}, 200 \mbox{
    (rectangles)}, 250 \mbox{ (diamonds)}$ and $300 \mbox{
    (down-triangles)}$ steps. (a) Plot of the specific
  energy (in $\varepsilon$-units) and (b) specific heat
  capacity (in $\varepsilon$-units) of the $8_1$ knot.}
\end{figure*}

\begin{table}[ht]
\caption{ Energy differences  between the unknot and the trefoil as
  well as those of
  the trefoil and 
  figure-eight. All knots have the same length. The values of the energy difference
  are reported in the case of various lengths $L=100,200,300$.
Here the temperature is $T=10\varepsilon$.}
\centering
\begin{tabular}{c c c c c c}
\hline\hline
 $L$ & unknot&trefoil&figure-eight&$d_{\cal UT}$&$d_{\cal TF}$\\ [0.5ex]
\hline
100 &0.174   &0.247   &0.279   &0.073   &0.032  \\
200 &0.170   &0.197   &0.210   &0.027   &0.013  \\
300 &0.167   &0.182   &0.186   &0.015   &0.004 \\ [1ex]
\hline
\end{tabular}
\label{table:nonlin}
\end{table}
To get a feeling of the size of the involved quantities, we can make
the concrete 
example of the proteins which,
as mentioned before, can be found in different topological
conformations like the 
trefoil and the figure-eight. These
proteins have
a length of about 250 residues \cite{dna} and their temperature
is around $310K$ (since this is approximately the temperature of the
human body). The parameter $\varepsilon$ for a protein is about 
$10K/mole$~\cite{ljwell}. Based on those data, for one mole, we can
roughly estimate 
the range of corresponding temperatures in our units, which is
$T/\varepsilon=310K/\varepsilon=31$. This temperature is located in the region
$T>T_{2}$.    
Before concluding this section, we would like to stress that the used
method for preserving the knot topology after a pivot move is very
time efficient and can be applied to any knot configuration
independently of its complexity. For example, in fig.~\ref{geo4} we
provide the data for the specific energy and heat capacity in the case
of the knot $8_1$. As it is possible to check, also for this knot the
same conclusions drawn before for ${\cal K}={\cal T},{\cal F}$ are
valid. In particular, 
the topological effects become less important with increasing polymer length
and it is confirmed that both the energy and
the heat capacity increase with increasing knot complexity.
\section{Conclusions}

We have studied the topological effects on the thermal properties of
several knot configurations with the help of  Monte Carlo simulations based on
the Wang-Landau algorithm and the pivot method. 
The topology of the knots, that can change after a pivot move, has been
preserved  by explicitly checking if during the move some parts of the
trajectory of
the knot have crossed themselves at some points.

 The
specific energy and the heat capacity of the trefoil, the figure-eight
and the $8_1$ knot
have been calculated for 
various lengths ranging from $100$ to $300$. 
As a test of our methodology,
these quantities have been derived also in the known case of the
unknot. In particular, we have been able to reproduce the
results of \cite{pnring}.
Our investigations show that the
specific energy and heat capacity 
for a polymer knot whose monomers are subjected to short-range
repulsive interactions
are depending on the length and the
topological conformation of the polymer knot. 
This could be expected because in the presence of repulsive
interactions the polymer will prefer ``diluted'' conformations
in which the monomer density is lower.
We may thus suggest
that the dilution of polymers in a good solvent should be effective in
reducing the 
topological effects. 
Moreover, for the growth of the specific energy with respect to the
temperature it is possible to
distinguish three different regimes that have been
explained in details in Section II. 
The zero energy regime appears at extremely low temperatures and it is
characterised by the fact that the energy of
the thermal fluctuations is below the threshold of the repulsive
energy barrier, so that the monomers have not enough energy to get
near.
After this threshold has been passed, the number of contacts between the
monomers starts to increase rapidly until at the end a saturation
phenomenon occurs and the energy of the system increases only slowly
with increasing temperatures. 
The plot of the specific heat capacity shows a pseudo-phase
transition which is related to the lattice geometry, as explained in
refs.~\cite{wustlandau} and \cite{vogel}.

Our simulations can be generalised under several aspects.
First of all,  we limited ourselves to simple short range
interactions, but there are no obstacles to extend our
procedure to describe more realistic polymer systems.
Moreover, we have limited ourselves to pivot moves involving a short
number of segments. No difference between the case $N=4$ and $N=5$
have been detected, so that all displayed diagrams are related only to the
case $N=5$. The independence of $N$ up to $N=10$ has been confirmed by
preliminary calculations  performed using another method
to preserve the topology based on knot invariants.
We hope to be able to report on these new
developments very soon \cite{zhaoferrari}.\\[1cm]

\begin{acknowledgments} 
We are indebted with V. G. Rostiashvili for helpful discussions and
suggestions 
that stimulated the development of
the method for distinguishing the topology of the different
knot configurations presented in this work. We wish also to thank
heartily J. Paturej and T. A. Vilgis for fruitful discussions.
The support of the Polish National Center of Science,
scientific project No. N~N202~326240, is gratefully acknowledged.
The simulations reported in this work were performed in part using the HPC
cluster HAL9000 of the Computing Centre of the Faculty of Mathematics
and Physics at the University of Szczecin.
\end{acknowledgments}

\end{document}